\documentclass[twocolumn,aps,prl,psfig,showpacs]{revtex4}
\usepackage{amsfonts}
\usepackage{amsmath}
\usepackage{graphicx}
\usepackage{bm}
\usepackage{amssymb}
\usepackage{times}
\usepackage{dcolumn}

\makeatletter

\begin{document}

\title{Face-Centered-Cubic B$_{80}$ Metal: Density functional theory calculations}
\author{Qing-Bo Yan}
\author{Qing-Rong Zheng}
\author{Gang Su}
\email[Author to whom correspondence should be addressed. ]{
Email:gsu@gucas.ac.cn}
\affiliation{{College of Physical Sciences, Graduate University of Chinese Academy of
Sciences, P.O. Box 4588, Beijing 100049, China}}

\begin{abstract}
By means of \emph{ab initio} calculations within the density functional
theory, we have found that B$_{80}$ fullerenes can condense to form stable
face-centered-cubic (fcc) solids. It is shown that when forming a crystal, B$%
_{80}$ cages are geometrically distorted, the I$_{h}$ symmetry is
lowered to T$_{h}$, and four boron-boron chemical bonds are formed
between every two nearest neighbor B$_{80}$ cages. The cohesive
energy of B$_{80}$ fcc solid is 0.23 eV/atom with respect to the
isolated B$_{80}$ fullerene. The calculated electronic structure
reveals that the fcc B$_{80}$ solid is a metal. The predicted solid
phase would constitute a new form of pure boron, and might have
diverse implications. In addition, a simple electron counting rule
is proposed, which could explain the stability of B$_{80}$ fullerene
and the recently predicted stable boron sheet.
\end{abstract}

\maketitle

Boron and carbon, the neighbors in the periodic table, both possess very
rich physical and chemical properties. Carbon forms the backbone of life, as
there are millions organic compounds that contain carbon in nature. Besides
the graphite and diamond, carbon also possesses new isomorphics, such as
carbon nanotube and fullerenes \cite{c60discover,fullerene_book}, which have
been extensively studied, and may be the fundamental materials for molecular
electronics. Boron has a variety of complex isomorphic structures, such as $%
\alpha $-rhombohedral B$_{12}$, $\beta $-rhombohedral B$_{105}$, and
tetragonal B$_{50}$, etc (see e.g. \cite{boronreview} for review). The boron
solids are usually band insulators or semiconductors \cite%
{borongap1,borongap2,borongap3}, but they could become metals or
superconductors under high pressure \cite%
{boronpressure1,boronpressure2,boronpressure3}. In the past decade, boron
sheet, cluster and boron nanotube (BNT) also gain wide attention \cite{BNT1,
BNT2, BNT3, BNT4, BNT5}, most of which were shown to be metallic, and they
are expected to have broad applications in various circumstances.

Recently, a new allotropic form of boron, B$_{80}$, with a round,
monoelemental, and hollow structure, has been predicted \cite{B80predict},
which is coined as boron fullerene. It consists of 80 boron atoms, and is
very similar in shape to C$_{60}$ fullerene except that an additional boron
atom sits in the center of each hexagon (Fig. 1(a)). B$_{80}$ could be one
of the most stable boron cages so far \cite{B180}, which, if confirmed
experimentally, should be the second example of a monoelemental buckyball
after C$_{60}$ in nature. Our further analyses indicate that B$_{80}$
fullerene is probably formed by multi-center deficient-electron bonds, and
satisfies a simple electron counting rule. This rule could also apply to
explain the outstanding stability of the new class of boron sheets and
nanotubes composed of triangles and hexagons predicted recently \cite%
{newboronsheet,newBNT}. On the other hand, recall that soon after discovery
of C$_{60}$, people found that C$_{60}$ clusters can condense to form solid
phases such as simple cubic (sc), face-centered-cubic (fcc) \cite{C60SCFCC},
and hexagonal-close-packed (hcp) \cite{C60HCP} crystals. As B$_{80}$ cluster
has a geometrical structure similar to C$_{60}$, and boron and carbon share
some similarities in chemistry \cite{BCsimilarity}, analogously, it would be
natural to anticipate that the boron fullerenes could also condense to form
solid phases. Here we show that, based on the \emph{ab initio} calculations
within the density functional theory (DFT) \cite{Hohenberg-Kohn}, B$_{80}$
clusters can indeed condense to form a stable fcc solid that has a metallic
electronic structure, in contrast to either the popular solid phases of pure
boron that are usually insulating or semiconducting, or the fcc and hcp C$%
_{60}$ that are band insulators \cite{c60solid_gap}. In addition, we find
that, unlike the case of C$_{60}$ where the cage structure retains intact
and is condensed by van der Walls force in solids\cite{C60SCFCC,
cohesive_solid_C60}, when B$_{80}$ clusters condense to form the solid
phase, the geometrical structure of B$_{80}$ is strikingly distorted [Fig.
1(d)], which is resulted from the formation of four boron-boron chemical
bonds between every two nearest neighbor B$_{80}$ clusters. The total energy
of B$_{80}$ fcc solid is 0.23 eV/atom lower than the isolated B$_{80}$
fullerene, and 0.35 eV/atom higher than $\alpha $-rhombohedral B$_{12}$
solid. Therefore, if the fcc B$_{80}$ metal is eventually confirmed
experimentally, it would compose a new form of pure boron in nature.

Our calculations are mainly performed by means of the ABINIT package \cite%
{ABINIT}. This package is coded within the DFT framework based on
pseudopotentials and plane waves, which relies on an efficient fast Fourier
transform algorithm \cite{FFT} for the conversion of wave functions between
real and reciprocal spaces, on the adaptation to a fixed potential of the
band-by-band conjugate gradient method \cite{Conjugate Gradient}, and on a
potential-based conjugate-gradient algorithm for the determination of the
self-consistent potential \cite{Gonze1,Gonze2,Gonze3}. Troullier-Martins
norm conserving pseudopotentials \cite{pseudo potential} generated by
fhi98PP code \cite{fhi98PP} are applied to mimic the electron-ion
interaction, and the Perdew-Wang 92 \cite{PW92} exchange-correlation
potential within local density approximation (LDA) \cite{Kohn-Sham} is used.
The kinetic energy cutoff in the plane-wave basis is taken as 25 Hartree,
and the tolerance for absolute differences of the total energy is set as 10$%
^{-6}$ Hartree.

\begin{figure}[tbp]
\includegraphics[width=0.90\linewidth,clip]{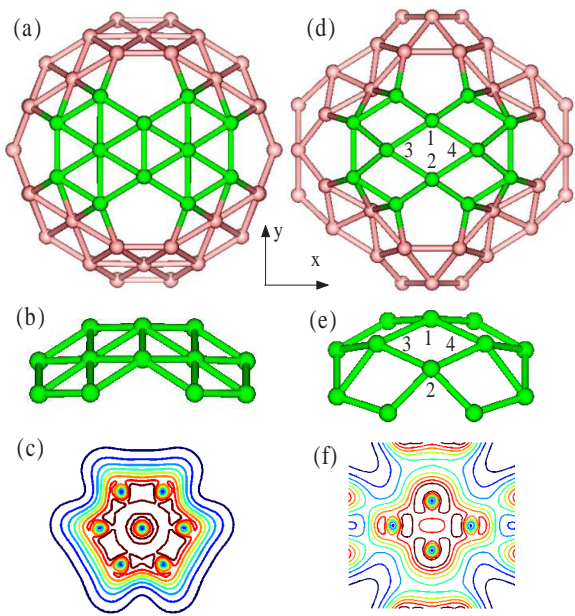}
\caption{(Color online) (a) Schematic structure of an isolated B$_{80}$
cluster; (b) is the view of green (dark) parts (a pair of neighbor hexagons)
in (a) after a $\protect\pi/6$ rotation around the \textit{x} axis; (c) the
contour map of valence electron density in the plane of a hexagon in (b);
(d) the distorted structure of B$_{80}$ in the fcc solid; (e) is the view of
green (dark) parts in (d) after a $\protect\pi/6$ rotation around the
\textit{x} axis, showing the distortion of a pair of neighbor hexagons in
(b); (f) the contour map of valence electron density in the plane of the
rhombus, which is composed of the atoms marked by number 1-4 in (e). The red
and blue colors in (c) and (f) represent high and low electron densities,
respectively. (See the color bar in Fig. 2.)}
\label{B80_structure}
\end{figure}

The structure of the isolated B$_{80}$ cluster is optimized in a cubic
supercell with a lattice parameter 25 $\mathring{A}$ by the
Broyden-Fletcher-Goldfarb-Shanno (BFGS) minimization \cite{BFGS} with a
convergence tolerance of 10$^{-5}$ Hartree/Bohr. As shown in Fig. 1(a), B$%
_{80}$ cluster, bearing I$_{h}$ symmetry, is similar in shape to C$_{60}$,
but there is an additional B atom centering in each hexagon, in agreement
with the early calculation \cite{B80predict}. Apparently, a pure two-center
two-electron (2c2e) bonding scheme could not give a proper description of B$%
_{80}$ fullerene, as in which each boron atom has five or six nearest
neighbors but it only possesses three valence electrons. We believe that the
picture of the multi-center deficient-electron bonding \cite{mcde} may apply
to the present case. One may note that a hexagon with an additional boron
atom sitting in the center can be viewed as a group of six triangles. A B$%
_{80}$ fullerene consists of 20 such hexagons and 12 pentagon holes, or
equivalently, 120 triangles and 12 pentagon holes. Suppose that three boron
atoms on the vertices of a triangle share a three-center two-electron (3c2e)
bond. Then, a triangle would consume two electrons. As there are 120
triangles in B$_{80}$, 240 electrons are demanded for bondings. While the
boron atoms on the vertices of pentagon holes have been counted in the
triangles, no electron is consumed for pentagon holes. Thus, the total
number of demanded electrons is 240, which is just equal to the total number
of electrons that B$_{80}$ possesses. Based on the multi-center
deficient-electron bonding picture, such a simple electron counting method
may be generalized as follows: in a stable boron quasiplanar structure
(sheet, nanotube, fullerene, etc.) composed by triangles and pentagon
(hexagon) holes, if we assume that each triangle consumes two electrons,
while pentagon (hexagon) holes consume no electrons, we would find that the
number of total electrons demanded for bondings, that should be equal to the
total number of valence electrons of the system, is nothing but twice the
number of triangles. This electron counting rule also works for the newly
predicted most stable boron sheet and nanotube\cite{newboronsheet,newBNT},
which are composed by triangles and hexagon holes. In those cases, 8 boron
atoms form a unit cell, which possess 12 triangles, so according to the
above rule, 24 electrons in total are required for bondings, which just
corresponds to the 24 valence electrons of 8 boron atoms, giving rise to a
bonding balance. As illustrated in Ref. \cite{newboronsheet}, a pure
hexagonal boron sheet is prone to accepting electrons, while a pure
triangular boron sheet has a surplus of electrons, leading to that a mixture
of these two phases with a proper proportion of triangles and hexagons would
make the structure more stable. Our electron counting rule may be understood
in a similar way. In a boron quasiplanar structure composed by triangles, a
hexagon (pentagon) hole could be produced by removing the central atom from
a group of six (five) triangles that share the central atom. As discussed
above, according to the multi-center deficient-electron bonding scheme, each
triangle consumes two electrons, such a removal will reduce one boron atom
that has three valence electrons, and therefore twelve (ten) electrons for
bondings are cut down simultaneously. In this way, one may change the number
of hexagon (pentagon) holes to realize the adjustment of the balance of the
supplied and demanded electrons. Consequently, the electron counting rule
manifests itself that a proper proportion of hexagon (pentagon) holes and
triangles should be kept to give rise to the most stable boron quasiplanar
structures. For instance, in B$_{80}$ cluster, the proportion of pentagon
holes and triangles is 1/10. Fig. 1(c) shows the valence electronic density
of the hexagons in B$_{80}$ cluster. One may see that high electron
densities exist around the center boron atoms, showing a ring-like
multi-center bonding pattern. As a hexagon with central atom [Fig. 1(c)] can
be viewed as a combination of six triangles, the ring-like multi-center bond
may be evolved from the combination of six 3c2e bonds.

To seek for stable solid phases of B$_{80}$, we have attempted sc and fcc
crystal structures, and taken two steps to find the optimized structures. In
the first step, with the atomic positions in B$_{80}$ clusters fixed, the
total energy of B$_{80}$ solid with the presumed lattice, was calculated in
every 0.1 $\mathnormal{\mathring{A}}$ for the lattice constant (where the
data with precision of 0.01 $\mathring{A}$ are obtained by a spline
interpolation from the calculated data with precision of 0.1 $\mathring{A}$%
). Without loss of generality, in the calculations a properly fixed
orientation of B$_{80}$ molecules is kept, namely, the $xy$ plane of B$_{80}$
molecule is arranged to be along the (100) plane in the sc and fcc
structures, which makes the system bear a relatively high symmetry. In order
to examine the effect of orientational disorder on the calculated results,
we have rotated B$_{80}$ clusters randomly to reduce the symmetrical
orientations, and recalculated the optimized lattice parameters and the
corresponding total energy. The results show that the total energy increase
a lot (several eV per B$_{80}$ cluster), suggesting that our presumed
geometrical configuration is energetically more favorable. We therefore get
the total energy as a function of lattice parameters for both sc and fcc
structures, where a single minimum in the curve of energy versus the lattice
constant is found, indicating that a stable structure may exist. In the
second step, on the basis of the structures obtained in the first step, a
full relaxation including the atomic positions, cell shape and volume was
conducted by means of the BFGS minimization until the forces acting on atoms
are less than a tolerance of 10$^{-5}$ Hartree/Bohr. As a result, after
relaxation, the shape, lattice parameter, and the total energy are all
varied, with the cohesive energies 4.9 eV and 18.2 eV per B$_{80}$ unit,
i.e., 0.06 and 0.23 eV/atom, for the sc and fcc B$_{80}$ solids,
respectively, which are measured with respect to the isolated B$_{80}$
cluster. As the fcc phase appears to be more stable in energy than the sc
phase, we will focus on the fcc phase in the following.

As shown in Fig. 1(d), B$_{80}$ cage is dramatically distorted in fcc solid,
where the I$_{h}$ symmetry of an isolated B$_{80}$ is lowered to T$_{h}$
symmetry for B$_{80}$ in fcc phase. To check it further, we have relaxed a
single distorted B$_{80}$ cage in a supercell, and found that it can indeed
revert to the isolated B$_{80}$ cluster [Fig. 1(a)] with a spherical shape.
Therefore, the distortion of B$_{80}$ cages in a fcc solid may be mainly
owing to the formation of boron-boron chemical bonds between nearest
neighbor B$_{80}$ cages. Fig. 1(b) and (e) show the configurations of a pair
of hexagons along the \emph{z} axis before and after distortion,
respectively. The sharing arris of the two neighboring hexagons is elongated
(from 1.678 $\mathring{A}$ to 1.940 $\mathring{A}$) and broken, while two
central atoms in the hexagons move outward, resulting in a planar rhombus
structure composed by atoms denoted by the number 1-4 in Fig. 1(e), where
the side length is 1.773 $\mathring{A}$ and the angle is 57.87$^{\circ }$.
The rhombus is surrounded by four skew quadrilaterals, two triangles and two
pentagons. Due to the symmetry, there are total six such pairs of hexagons
along the \textit{x}, \textit{y} and \textit{z} axes in one isolated B$_{80}$
cluster, which evolve into a deformed B$_{80}$ unit with six geometrically
equivalent rhombuses. The boron atoms sitting in the centers of the rest
eight hexagons in a B$_{80}$ unit all shift inward. Fig. 1(f) presents the
valence electron density of the rhombus structure. From the profile, we
observe that the boron atoms labeled by 3, 1, 4 and 3, 2, 4 form two B-B-B
bridge-type 3c2e bonds, which are very similar to the B-H-B bridge-type 3c2e
bond in B$_{2}$H$_{6}$. The valence electron density is rather high between
every two of the four boron atoms, and an elliptical high-density area
presents in the center of the rhombus. It appears that the two B-B-B
bridge-type 3c2e bonds are coupled with each other, which eventually become
a new complicated four-center bond. It will be illustrated later that the
formation of this four-center rhombus is crucial to the bondings between the
nearest neighbor B$_{80}$ units.

\begin{figure}[tbp]
\includegraphics[width=1.0\linewidth,clip]{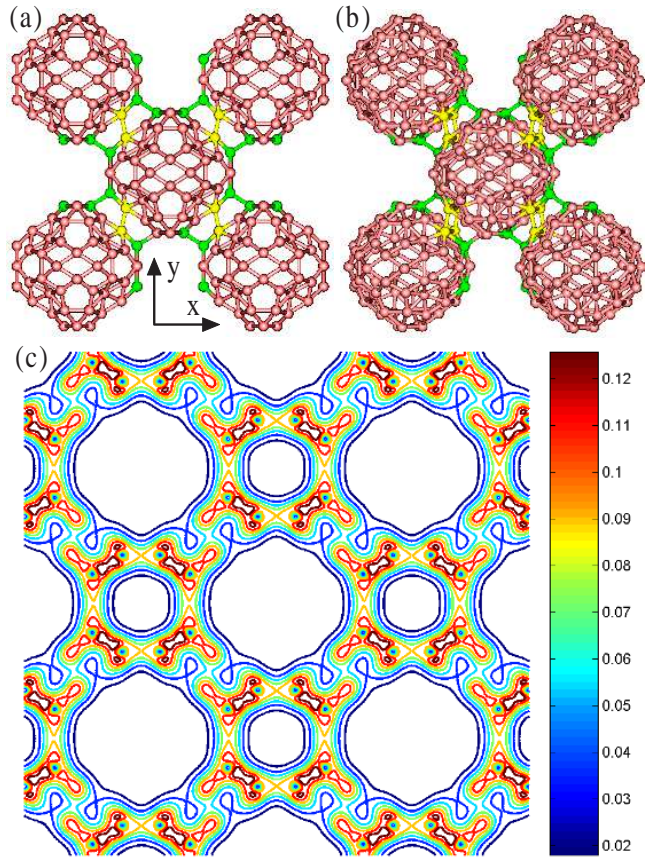}
\caption{(Color online) (a) Schematic map of the bondings of the nearest
neighboring five B$_{80}$ clusters in (001) plane, the lengths of the green
(dark) and yellow (grey) bonds are 1.734 $\mathring{A}$ and 1.700 $\mathring{%
A}$, respectively; (b) a slightly rotational view of (a), where four
boron-boron chemical bonds between the nearest neighbor B$_{80}$ units can
be seen clearly; (c) the contour map of valence electron density of the fcc B%
$_{80}$ solid (a) in the (001) plane through the center of B$_{80}$ units,
corresponding to the cross section of (a). The red and blue color represent
high and low electron densities, respectively. The unit of color bar is
electrons/Bohr$^{3}$.}
\end{figure}

As can be seen in Figs. 2(a) and (b), there are four boron-boron chemical
bonds between every two nearest neighbor B$_{80}$ units. The two green
(dark) bonds are formed between two vertical rhombuses, while the two yellow
(grey) bonds connect two parallel pentagons. The lengths of the green and
yellow bonds are 1.734 $\mathring{A}$ and 1.700 $\mathring{A}$,
respectively, both falling into the typical boron-boron bond length scales
in B$_{80}$ fullerene \cite{B80predict} and other boron materials. In a fcc B%
$_{80}$ solid, every B$_{80}$ unit has 12 nearest neighbor B$_{80}$ units,
and every two nearest neighbor B$_{80}$ units are bonded by four boron-boron
chemical bonds, 48 bonds are therefore formed between every B$_{80}$ unit
and its 12 neighbors. In this sense, the fcc B$_{80}$ solid can be regarded
as a three-dimension network of the distorted B$_{80}$ units. Fig. 2(c)
presents the valence electron densities of the fcc B$_{80}$ solid in the
(001) plane through the center of B$_{80}$ units. The five big circles
indicate the inner hollows of the five B$_{80}$ units in Fig. 2(a). It can
be observed that there are high electron densities between neighboring
cages, indicating stronger interactions between neighboring B$_{80}$ units.
In particular, the densities corresponding to the green bonds are rather
high, showing a $\sigma $-like bond character. (The yellow bonds do not lay
in this plane and are not shown directly in the density map.) Thus, we may
conclude that B$_{80}$ clusters in the fcc phase are connected by strong B-B
$\sigma $-like chemical bonds.

\begin{figure}[tp]
\includegraphics[width=1.0\linewidth,clip]{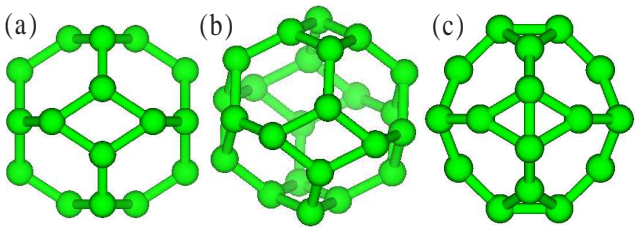}
\caption{(Color online) (a) The cage-like B$_{24}$ structure between the
neighbor B$_{80}$ units; (b) is the rotational view of (a); (c) is the
distorted structure after relaxation in a supercell, where the rhombus
breaks into two triangles.}
\end{figure}

In Fig. 2(c), it seems that small spherical-like hollow structures are
formed in the space between the neighbor B$_{80}$ units. By looking into it
further, we find that it is indeed a B$_{24}$ cage-like structure composed
of 6 rhombuses and 8 hexagons [Fig. 3(a) and (b)]. This structure could be
considered as a distorted truncated octahedron or cubo-octahedron, and
interestingly, it is also a regular part of the crystal structure of
metal-dodecaborites MB$_{12}$ (M is a metal) \cite{MB12}, in which a metal
atom is located at the center of each boron cubo-octahedron. We have also
relaxed the B$_{24}$ cage in a supercell, and seen that it is distorted and
becomes more spherical, and every rhombus breaks into two triangles [Fig.
3(c)]. The total energy of the relaxed B$_{24}$ cage is 0.57 eV/atom larger
than the isolated B$_{80}$ cluster. Thus, this B$_{24}$ cubo-octahedral
structure could exist only in a solid, and cannot appear in an isolated
form. Similar to the situation of MB$_{12}$, the interstitial atoms may be
inserted into these hollows to form new boron compounds.

\begin{table}[bp]
\caption{Total energies of the boron sheet, isolated B$_{80}$ cluster, and B$%
_{80}$ fcc solid, with respect to $\protect\alpha$-rhomhedral boron solid,
in eV/atom. }
\label{Table1}\setlength{\textfloatsep}{0.5cm}
\begin{tabular}{cccc}
\hline\hline
& boron sheet & isolated B$_{80}$ & B$_{80}$ fcc solid \\ \hline
Ref. \cite{newboronsheet} & 0.38 & 0.56 & - \\
This work & 0.46 & 0.58 & 0.35 \\ \hline\hline
\end{tabular}%
\end{table}

As listed in Table I, the total energy of fcc B$_{80}$ solid is 0.23 eV/atom
lower than the isolated B$_{80}$, 0.35 eV/atom higher than $\alpha$%
-rhombohedral B$_{12}$ solid. Note that the cohesive energy of the fcc C$%
_{60}$ solid is about 1.6 eV per C$_{60}$ \cite{cohesive_solid_C60}, i.e.,
0.03 eV/atom. Furthermore, fcc C$_{60}$ solid and other molecular solids
\cite{yan,yan1,c60solid_narrowbands} are condensed by van der Walls force,
in which no chemical bonds are formed among the neighboring clusters. Our
calculations on the inter-cluster bonds, valence electron densities, and the
cohesive energies indicate that the condensed mechanism of fcc B$_{80}$
solid is totally different from the above molecular solids.

The energy bands and the density of states (DOS) for the fcc B$_{80}$ solid
are calculated with the optimized lattice constants, as presented in Fig. 4.
The energy bands are quite dispersive, and several bands spread across the
Fermi level, as manifested in the left panel of Fig. 4, showing that the fcc
B$_{80}$ solid is a metal. As the most of the BNTs are also observed to have
the metallic electronic structures, such a metallic property of the fcc B$%
_{80}$ solid is conceivable. Unlike the fcc C$_{60}$ solid that is a band
insulator with a direct energy gap of 1.5 $eV$ \cite{cohesive_solid_C60},
the present result shows that the energy gap closes when B$_{80}$ clusters
condense to form a fcc solid. This is understandable, because there are
strong interactions between the neighboring B$_{80}$ clusters in a solid
phase, which would inevitably enhance the overlap of the charge densities of
electrons, thus broadening the energy bands, and eventually leading to the
vanishing of the energy gap. However, albeit the fcc B$_{80}$ solid shows a
metallic behavior, the electronic density at the Fermi surface is not so
high, as shown in the right panel of Fig. 4, in which the DOS profile is
depicted. The DOS has a minimum slightly above the Fermi level, giving rise
to a low electronic density of states. Of particular interest is that the
DOS profile of the fcc B$_{80}$ around the Fermi level look very similar to
that of MgB$_{2}$, in which the DOS of MgB$_{2}$ also exhibits a minimum
just above the Fermi level \cite{singh,Kortus}. MgB$_{2}$ is a well-known
two-gap superconductor with transition temperature T$_{c}$ $\sim$ 39 K \cite%
{Nagamatsu}, where it has been established that the energy bands at the
Fermi level mainly derive from B orbitals, and is a typical metal that is
essentially attributed to metallic boron with covalent B-B and ionic B-Mg
bonding \cite{Kortus}, whose superconductivity is now well understood within
the electron-phonon mechanism. This present analogy between the primary
characters of electronic structures of fcc B$_{80}$ and MgB$_{2}$ gives a
hint that superconductivity might exist in the fcc B$_{80}$ crystal.

\begin{figure}[tbp]
\includegraphics[width=0.80\linewidth,clip]{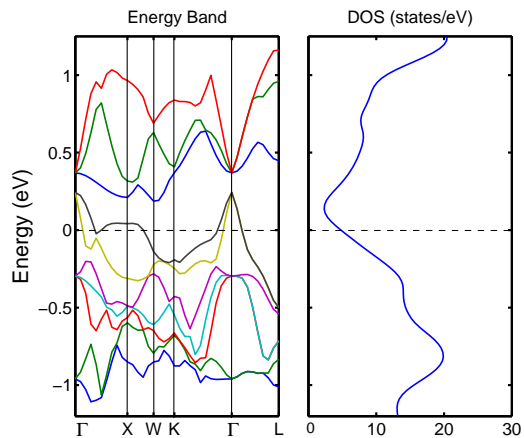}
\caption{(Color online) Left panel: Energy bands of the fcc B$_{80}$
crystal. Right panel: Density of states (DOS) of electrons for the fcc B$%
_{80}$ crystal around the Fermi level, which is obtained by means of a
smearing technique. The Fermi level is set to zero.}
\label{B80_band_dos}
\end{figure}

In conclusion, in terms of the first-principles DFT calculations, we
have found that B$_{80}$ fullerenes can condense to form stable sc
and fcc solids. It is also uncovered that B$_{80}$ cages in fcc
solid phase are geometrically distorted, where the I$_{h}$ symmetry
is lowered to T$_{h}$, and four boron-boron chemical bonds are
formed between every two nearest neighbor B$_{80}$ cages. The total
energy of B$_{80}$ fcc solid is 0.23 eV/atom lower than the isolated
B$_{80}$ fullerene. A simple electron counting rule is proposed,
which could explain the stability of B$_{80}$ fullerene and the
recently predicted most stable boron sheet. In comparison to the
ordinary semiconducting boron crystals that can become a metal or
superconductor under high pressure, our calculated electronic
structures, with some primary features quite similar to those of
MgB$_{2}$ superconductor, show that the fcc B$_{80}$ solid is a
metal at ambient pressure, and may be a candidate of new
superconductor. As a result, in spite of the popular phases of
$\alpha $-, $\beta $- rhombohedral and tetragonal borons, the fcc
B$_{80}$ metal may be another novel form for pure boron in nature.

The authors are grateful to X. Chen, S. S. Gong, and Z. C. Wang for helpful
discussions. All of the calculations are completed on the supercomputer
NOVASCALE 6800 in Computer NetWork Information Center (Supercomputing
Center) of Chinese Academy of Sciences. This work is supported in part by
the National Science Foundation of China (Grant Nos. 90403036, 20490210),
the National Science Fund for Distinguished Young Scholars of China (Grant
No. 10625419), the MOST of China (Grant No. 2006CB601102), and the Chinese
Academy of Sciences.

\end{document}